# RecoMed: A Knowledge-Aware Recommender System for Hypertension Medications

Maryam Sajde[1],
Hamed Malek[2],
Mehran Mohsenzadeh[1]

[1] Computer Engineering Department, Science and Research Branch, Islamic Azad University, Tehran, Iran.
[2] Faculty of Computer Science and Engineering, Shahid Beheshti University, Tehran, Iran.

# Abstract

### Background & Objective

High medicine diversity has always been a significant challenge for prescription, causing confusion or doubt in physicians' decision-making process. This paper aims to develop a medicine recommender system called RecoMed to aid the physician in the prescription process of hypertension by providing information about what medications have been prescribed by other doctors and figuring out what other medicines can be recommended in addition to the one in question.

### Methods

There are two steps to the developed method: First, association rule mining algorithms are employed to find medicine association rules. The second step entails graph mining and clustering to present an enriched recommendation via ATC code, which itself comprises several steps. First, the initial graph is constructed from historical prescription data. Then, data pruning is performed in the second step, after which the medicines with a high repetition rate are removed at the discretion of a general medical practitioner. Next, the medicines are matched to a well-known medicine classification system called the ATC code to provide an enriched recommendation. And finally, the DBSCAN and Louvain algorithms cluster medicines in the final step.

### Results

A list of recommended medicines is provided as the system's output, and physicians can choose one or more of the medicines based on the patient's clinical symptoms. Only the medicines of class #2, related to high blood pressure medications, are used to assess the system's performance. The results obtained from this system have been reviewed and confirmed by an expert in this field.

**Keywords:** Medicine recommender systems, Healthcare system, Hypertension, ATC code

# 1. Introduction

With the increase in data volume, finding the user's desired items among the sea of data has become a relatively difficult task, resulting in user confusion or incorrect selection [1]–[4]. Recommender systems aim to analyze the user's behavior and recommend the most similar item to his preferences. Audiences have grown to appreciate recommender systems over time [5]–[10]. In recent years, health recommendation systems especially in medicine domain have become increasingly popular [11]–[13].

For example, in 2016, researchers developed a recommender system that could recommend an attending physician based on the patient's health profile. Users can search for patients who are similar to them and attending physicians. A similar issue has been investigated in recent years [14]–[16]. DIETOS (DIET Organizer System), a unique dietary recommender system for chronic diseases, including patients with chronic kidney disease (CKD), was introduced in additional studies published in 2016 and 2017. With adjusted and dynamic questionnaires developed by physicians based on the user's physical state and chronic conditions [17]–[19], this system recommends a healthy diet.

In another work, SousChef, created by a group of researchers, was introduced as a meal recommender system for adults and older people[20]. s

As a result of other research activities, the diabetes care system became operational in 2017. It provides users with services, such as potential medicines, diet, and appropriate workout, by comparing patients' records and finding similar users based on user characteristics [21]–[23].

In 2018, a recommender system was introduced by Chiang et al. to immunize medicine recommendations to prevent adverse medicine reactions (ADRs) due to medicine-medicine interaction (DDI) due to concomitant use of multiple medicines [24].

In other recently published articles, a medicine recommender system was also designed to reduce medicine error using the SVM algorithm and data mining techniques [25], [26].

Also, with the arrival of the coronavirus in 2020, people began to self-medicate (take medicines indiscriminately) as a result of a lack of access to reliable clinical and medical resources and a dearth of specialists and healthcare workers, and adequate equipment. Using classification algorithms, in a paper presented in 2021, a medicine recommender system was introduced to reduce the need for professional presence [27]. A personalized recommender system was also built utilizing image processing techniques in a study published in 2021 to rate and find the correct medicine for cancer cells [28].

In another work, an effective personalized recommendation of a travel recommender system has been developed [29]. It plays a vital role in reducing the time and cost of travel for travelers. Also, Conversational recommender systems have been designed by some researchers [30]. As a result, a vast number of people all over the world use these systems but it can be more suitable for people who live in deprived areas since the people who live in this place don't have easy access to the information and expert people and they face serious problems especially in healthcare.

According to WHO statistics and reports, cardiovascular diseases are one of the leading causes of death among men and women worldwide, especially in low-income countries. Indeed, hypertension is one of the severe problems that can increase the risk of heart disease, kidney, and other diseases. One out of four men and one out of five women suffer from hypertension disease [31]–[33]. Thus, the purpose and range of medicines studied in this study have been narrowed to high blood pressure medications for more practical application.

Proper prescription has always been challenging for physicians due to physician inexperience, medicine diversity on the market, and ignorance of recent developments, which can sometimes perplex or confuse physicians about prescription[34]. Recommender systems can be considered as a viable solution in these circumstances.

In this paper, a medicine recommender system called RecoMed is proposed to aid the physician in the prescription process of hypertension by providing information about what medications have been prescribed by other doctors and figuring out what other medicines can be prescribed in addition to the one in question. Data that is being used in this project was collected from a large number of prescriptions from 40 pharmacies all over Iran. The description of the dataset and how data gathering was done is covered in section 4. To reduce the medicine-medicine interaction errors and increase the confidence coefficient of the proposed system, and quality control of the recommended items, a standard medicine classification called ATC code, an anatomical therapeutic chemical classification system [35]–[39] was employed. By entering medicine data, the system should be able to recommend a suitable medicine.

The remainder of this paper is organized as follows: Section 2 outline the proposed method for resolving this problem. Section 3 presents the results. Section 4 is dedicated to comparing and evaluating the proposed method. And finally, Section 5 covers the conclusion and future work.

## 2. Methodology

Two steps are taken to facilitate the prescription process for the physician and achieve a suitable medicine recommendation when using the proposed approach and designing a medicine recommender system. The first step is to mine medicine association rules, followed by the second step in which graph mining is performed and an enriched recommendation is produced using the ATC code [35]–[38]. Doctors, particularly those with little experience practicing in underserved and remote areas, are among the target community of the system. This is due to fatigue, forgetfulness, inexperience, and an increase or lack of sure medicines on the market, to name a few factors.

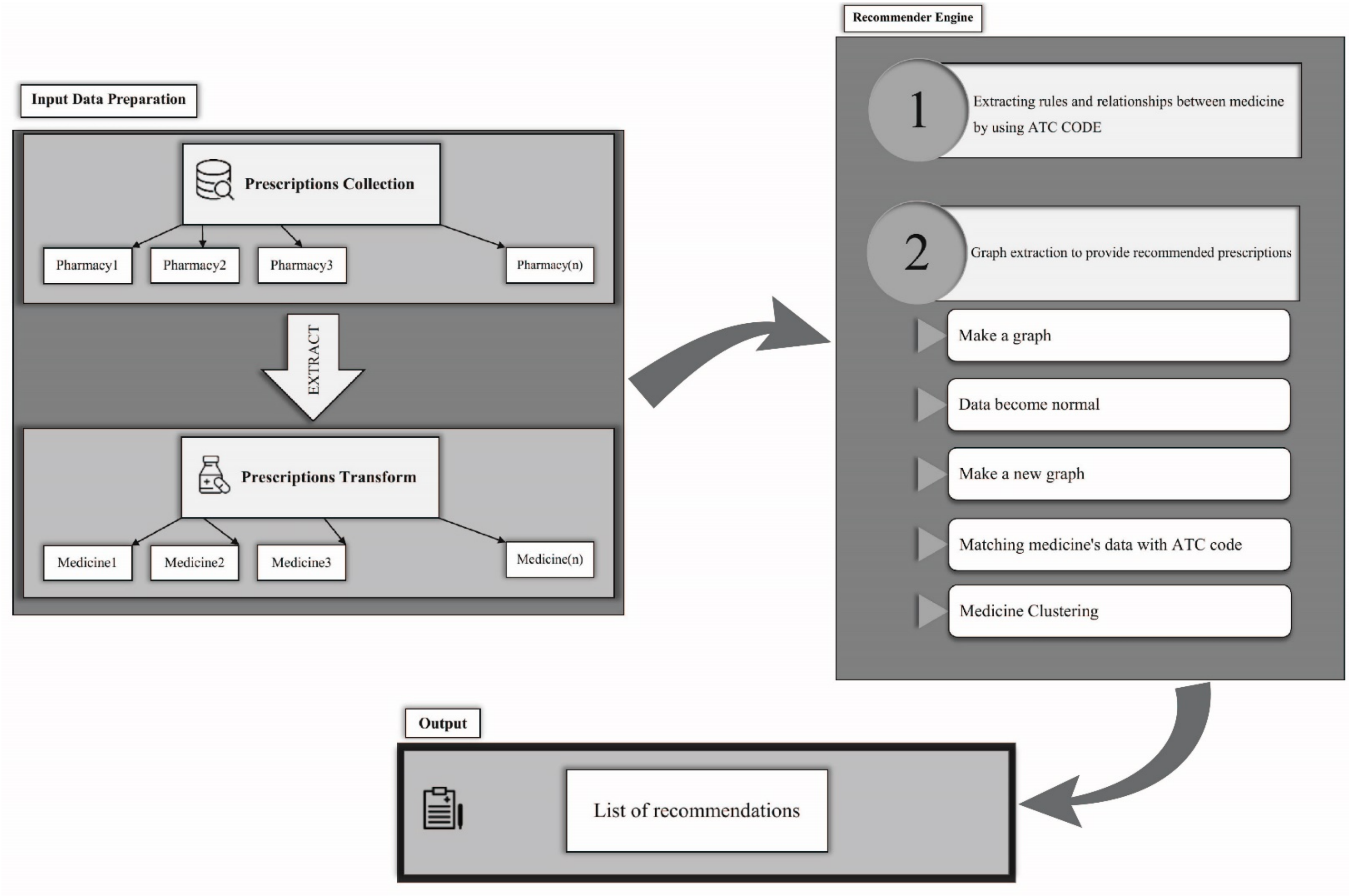

**Figure 1**: Medicine recommender system framework

## 2-1. The primary framework of the proposed recommender system

The proposed recommender system's macro framework is divided into three sections: (1) Input Data Preparation, (2) Recommender Engine, and (3) Output

**Input Data Preparation**

To prepare the input data, as shown in the diagram above, the input data preparation process is divided into two steps: (i) prescription collection and (ii) prescription transformation. A detailed description of the dataset is presented in Section 3.

**Recommender Engine**

The input data from the previous step is refined by discovering the rules between medicines using association rule mining algorithms [40]. Then, graph mining is used, and an enriched recommendation using the ATC code (a well-known medicine classification system) is produced. For graph mining, the initial graph is built with the nodes representing prescription medicines and the edges indicating the presence of two medications in the same prescription [41]. Then, data pruning operations are applied to remove general medicines with a high repetition rate from the database at the judgment of a medical expert. This results in a reconstructed graph where the unneeded medicines have been removed from the database. Besides, the medicines are adapted using a well-known medicine classification system, with the ATC codes of each medicine linked to the comparable medicine in the prescription. Finally, medicine clustering is done using the DBSCAN and Louvain algorithms detailed in Section 2-2.

**Output**

In summary, prescription medicines are individually entered into the system by physicians. After the system has processed the input data, the system output is a list of recommended medicines. The physician chooses one or more of the available recommended medicines based on the patient's clinical symptoms and then expresses their final opinion.

## 2-2. Steps to implement the recommender engine

### Step 1: Association rules

The association rules between medicines are mined using association rule mining algorithms in the first step. Since medicines can have many side effects on each other, in each group, the medicines with the highest interdependence should be selected, and the medicines with low incidence rates relative to one other should not fall into the class of recommended medicines because they can have adverse effects on each other.

In this work, a clustering method called Apriori Algorithm [34] which does not require labeled data is used. It is consistent with the datasets used in this study, which included prescriptions with unlabeled medicines. Furthermore, a large number of prescriptions are collected daily from all over the country. The dataset used in this project has 11 million records and it was collected from 40 pharmacies all over Iran. The description of the dataset and how data gathering was done is covered in section 4. Assuming that prescriptions must be processed at the end of each day and the data should be properly clustered adequately, a relatively lightweight and straightforward association rule mining algorithm is employed. The Apriori Algorithm refines the medicine data into prescription medicines in this step. Support, confidence, and lift are the three scalar numbers used to assess the medicine association rules discovered by this algorithm. This algorithm produces strong and weak rules. Which medicine pairs are most repetitive is determined by strong rules. On the other hand, a weak rule shows that as an example, these two particular medicines will have adverse effects and should not be prescribed with each other. The flowchart in Figure 2 depicts the association rule mining algorithm used in this work. A physician (or expert) determines the values of the min_sup and min_conf parameters by running a test and observing the results.

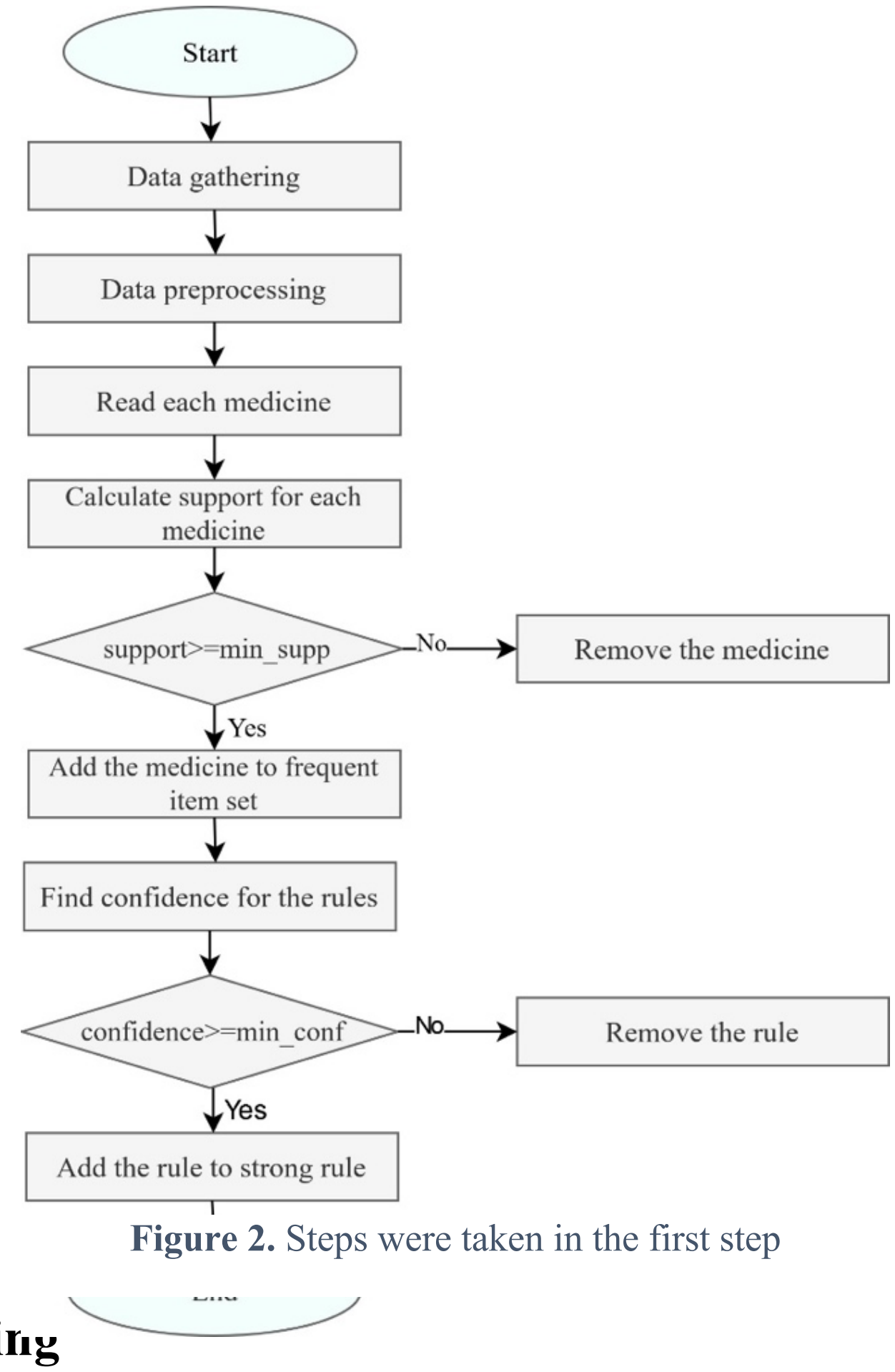

**Figure 2.** Steps were taken in the first step

## Step 2: Graph mining

Graph algorithms are powerful tools for demonstrating association and patterns between highly related datasets. Due to the complex nature of medical data and the close association between them, in this study, graph methods are used to determine the correct medicine classification and medicine association. More sensible recommendations can be made once the association between medicines is discovered. The details of the steps of the proposed method for constructing the final graph are as follows:

**Initial Graph**

The initial graph is built with the nodes representing prescription medicines and the edges indicating the presence of two medicines in the same prescription.

**Data Pruning**

According to medical studies, pharmacology books, and authoritative medicine authorities, some medicines are considered general in the medical world. They are not used only to treat a specific disease and have no side effects or medicine-medicine interactions. However, their recurrence in multiple prescriptions and association with other medicines provide little information. Hence, a table titled "Common Medicines" with several medicines is considered. The medicines in this table are named “Stop Medicines”. These medicines do not provide specific information; instead, it causes the graph to become cluttered. As a result of the medical experts' comments, it was decided to remove stop medicines from the graph. For this purpose, a graph pruning technique called the Natural Breaks algorithm, also known as the Fisher-Jenks algorithm, is used in this case [42], [43].

**Graph Rebuilding**

After deleting the list of stop medicines from the database, the graph must be rebuilt using the Jaccard Similarity algorithm [44] for clustering. The degree of similarity and diversity of data in a database is examined using this algorithm. This algorithm was chosen to compare the similarity of two medicines. The number of prescriptions in which the two medicines are present simultaneously determines the degree of similarity. The Jaccard coefficient is a number in the range of 0 to 1. The higher the coefficient, the two medicines exist in more prescriptions together.

**ATC Matching**

In the next step, Anatomical Therapeutic Chemical (ATC) classification system is used to match medicines with a well-known medicine classification system, check the accuracy of the recommendations, and produce a valid recommendation. The ATC code is the name given to the codes assigned to this system. ATC code divides medicines into 14 groups based on the organ or system they affect and their therapeutic, pharmacological, and chemical properties. The ATC codes of each medicine are then mapped to their equivalent medicine in prescriptions, and then clustering is applied in the next step. Medicines in the dataset are matched up with medicines in the ATC Code database peer-to-peer.

**Clustering**

In the final step, medicine clustering is performed using the DBSCAN and Louvain algorithms. The Louvain algorithm excels at data clustering and evaluating complex and large structures due to its one-of-a-kind nature [45]–[47]. However, while this algorithm can accurately cluster data, it cannot detect outliers. DBSCAN, on the other hand, has a solid ability to detect outlier data in addition to data clustering. But, on the other hand, it does not provide accurate clustering [4], [48]. Accurate clustering is critical in a medicine recommender system, as any error in medicine clustering results in bad consequences.

Hence, using the DBSCAN algorithm alone to implement this step does not produce a satisfactory result. Many factors, such as pharmacies in drugstores and insurance, have, on the other hand, play a role in the final medicines delivered in the collected dataset. As a result, errors and outlier data could exist in the existing database. Accordingly, both algorithms are being used at this point to compensate for each other's flaws. Outlier data is found using the DBSCAN algorithm first. And then, the Louvain algorithm is employed for clustering after the outlier data has been detected.

# 3. Evaluation

This section presents the outcomes of putting the proposed method into practice. A database with around 11 million records in JSON format from ElasticSearch has been provided to us for this study. The records of this dataset contain the dispensed prescriptions that have been collected from pharmacies all over the country. Medicine name, generic code of each medicine, medicine quantities, pharmacy name, and pharmacy location are among the fields extracted from these prescriptions for this project. The implementation of the system was done by Python and Scikit-learn library [49]. After trying and testing different sample sizes, it was found that 10.000 data is the minimum number of records for reaching the best result in this project. Therefore, a subset of the entire dataset consisting of 10,000 instances has been randomly selected to save time and facilitate the process of model building [41]. Finally, in order to evaluate the results of the implemented system, since unsupervised algorithms have been employed in this project, manual evaluation was performed by an expert person for verifying and confirming the outcomes.

## 3-1. Association rules

The output of this algorithm appeared as strong, and weak rules. Strong rules are used as indicators of which pairs of medicines are most frequently repeated together. On the contrary, by examining weak rules, it is determined which medicine will have an adverse effect on the event of another medicine, and do not appear together in a prescription. Therefore, they should not be recommended as a replacement. Strong and weak associations are identified according to support, confidence, and lift parameters.

Additionally, comparing the confidence value to the minimum confidence threshold (MCT) and the support value to the minimum support threshold (MST) can help to determine whether the achieved rule is strong or not. The values of MCT and MST values are set by an expert. For that, the MST value was set to 0.8 at first. Following that, more rules were discovered through the doctor's examinations. Hence, some medicines, such as serum and Dexamethasone, were involved in various rules and resulted in a large number of them. MST was then re-initialized, and based on the findings, physicians and experts decided on 0.9. A rule's strength or weakness can also be determined using the Lift parameter. Table 1 shows the output of this step.

## 3-2. Result of the second step

### Initial Graph

The graph before data pruning is depicted in Figure 3. Medicine data is divided into seven classes in this diagram. It represents the pre-pruning medicine graph based on prescription medicines extracted directly from the dataset.

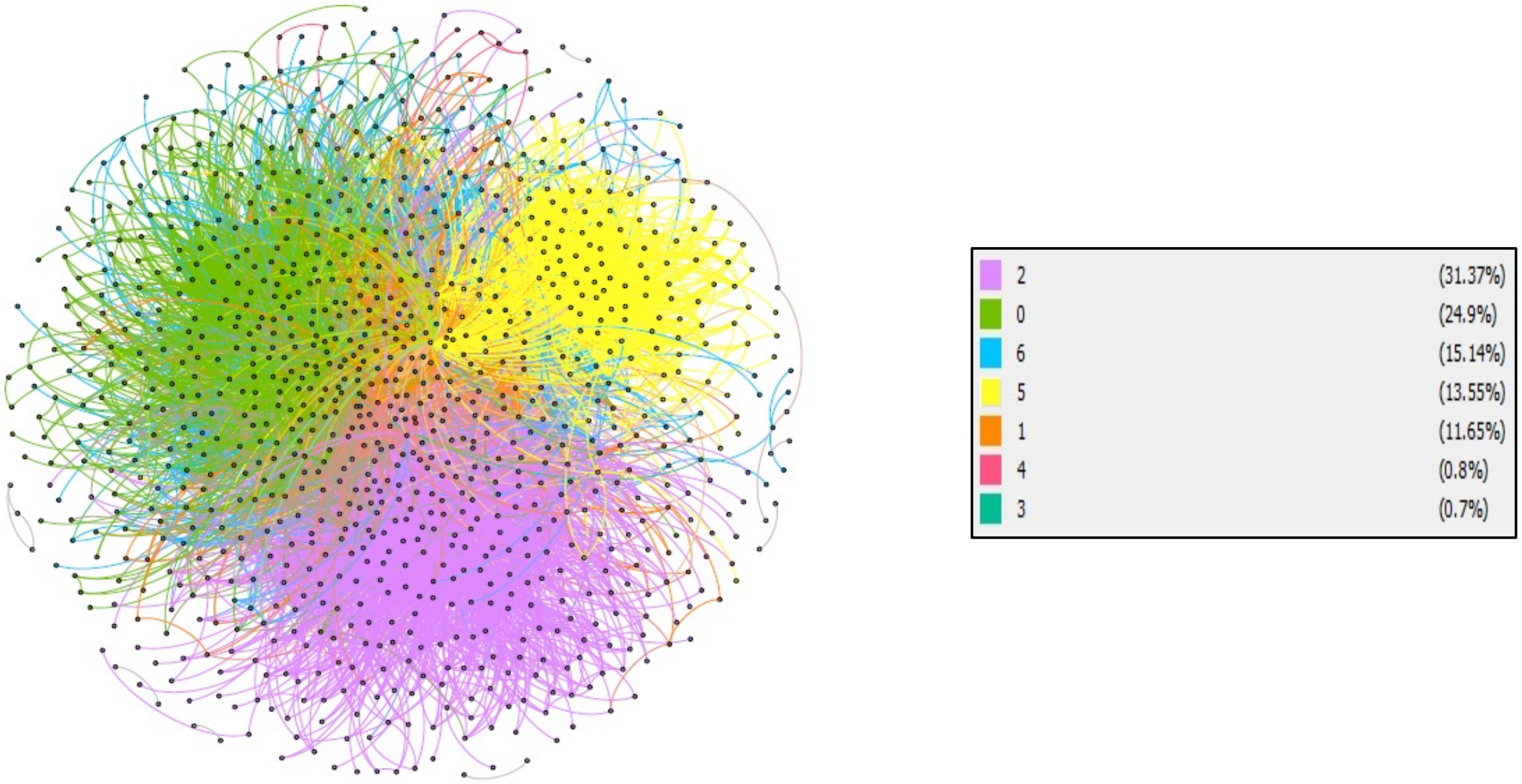

**Figure 3.** A graph is drawn based on pre-pruning prescription medicines

**Table 1.** The output of the Association Rule Mining algorithm

| Antecedents | Consequents | Support | Confidence | Lift |
|---|---|---|---|---|
| 1-PENICILLIN G PROCAINE 800,000 U VIAL<br>2-PENICILLIN G BENZATHINE (PEN LA) 1,200,000 U VIAL | WATER FOR INJECTION 5ML P-AMP | 0.0015 | 1 | 13.17 |
| 1- DOXORUBICIN HCL 10MG VIAL,<br>2- CYCLOPHOSPHAMIDE 500MG VIAL | APREPITANT 125/80/80MG CAP | 0.0013 | 1 | 66.86 |
| 1-WATER FOR INJECTION 5ML P-AMP,<br>2- CEFALEXIN 250MG/5ML 100ML POW FOR SUSP | PENICILLIN 6-3-3 VIAL | 0.0015 | 1 | 38.98 |
| 1-CIMETIDINE 200MG/2ML AMP,<br>2- GRANISETRON 3MG/3ML AMP | DEXAMETHASONE 8MG/2ML AMP | 0.0011 | 1 | 20.75 |
| 1-DOCETAXEL 20MG VIAL,<br>2- SET SERUM MEDI SMART | SODIUM CHLORIDE 0.9% 0.5L INF P-BOTTLE | 0.0016 | 1 | 14.67 |
| 1-FILGRASTIM(GCSF) 300MCG/ML INJ,<br>2- GEMCITABINE HCL 200MG VIAL APREPITANT 125/80/80MG CAP | GEMCITABINE HCL 1 G VIAL | 0.0011 | 1 | 26.74 |
| 1-CYCLOPHOSPHAMIDE 500MG VIAL,<br>2-DEXAMETHASONE 8MG/2ML AMP,<br>3- DOXORUBICIN HCL 50MG VIAL,<br>4-PEGFILGRASTIM 6MG/0.6ML INJECTION,<br>5- APREPITANT 125/80/80MG CAP | GRANISETRON 3MG/3ML AMP | 0.0014 | 1 | 52.04 |

## Data Pruning

The result of applying the Jenks Natural Breaks algorithm to the data is also shown in Figure 4 and Table 2.

**Table 2**. Fisher-Jenks algorithm output

| Cut_Jenks | Min | Max | Count |
|---|---|---|---|
| 0 | 1 | 44 | **1262** |
| 1 | 45 | 144 | **125** |
| 2 | 157 | 299 | **40** |
| 3 | 325 | 464 | **11** |
| 4 | 656 | 961 | **3** |

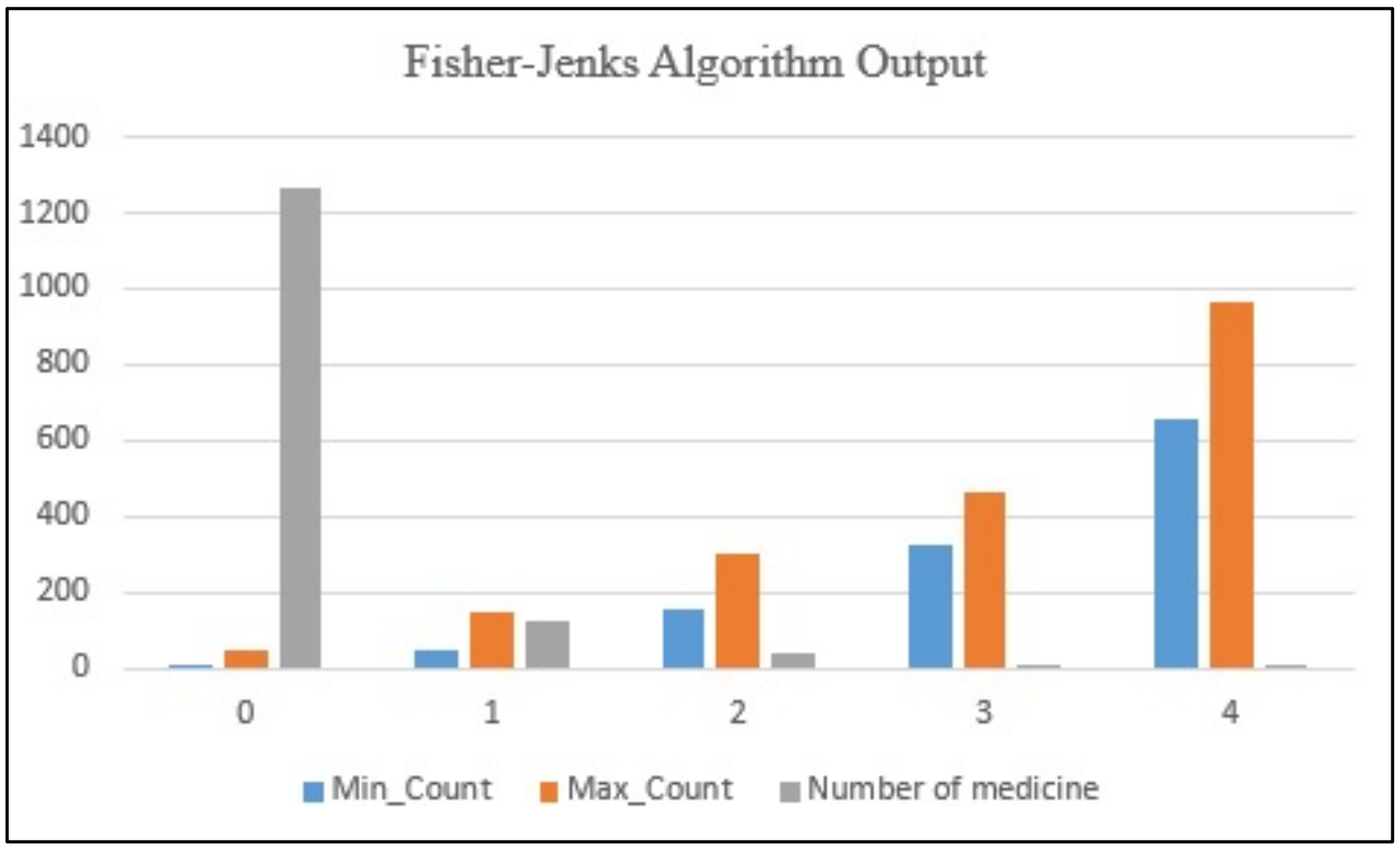

**Figure 4.** Medicine classification based on the Fisher-Jenks algorithm

According to the density of prescription medicines, the cut_jenks column is divided into clusters ranging from 0 to 4. The minimum and the maximum number of prescriptions in which medicine has been involved in each cluster are listed in the Min and Max columns, respectively. In addition, the number of medicines in each cluster are listed in the count column. In cluster 0, for example, Min = 1, Max = 44, and count = 1262. This means that each of the 1262 medicines in this cluster has at least one prescription and a maximum of 44.

By applying the pruning step, stop medicines are removed. By this, the graph's edges connected to the pruned nodes are also removed. This is referred to as graph pruning. The number of edges is reduced after these medicines were eliminated. To remove the medicines, the number of repetitions of existing medicines was used to divide them into quarters. The most commonly used medicines in this collection are Class 3 medicines, which appear in at least 656 and up to 961 prescriptions, and Class 4 medicines, which appear in at least 325 and up to 464 prescriptions. There are 14 medicines in total. As a result, the medicines in these two classes were removed from prescriptions with the doctor's approval by matching them to the table of stop medicines. The edges connecting these graph nodes were also removed after deleting these medicines. Table 3 shows the outcome of this removal operation.

**Table 3.** Number of edges and nodes after removal of repetitive medicine

| Parameters | Before removing 14 medicines | After removing 14 medicines |
|---|---|---|
| Number of edges | 16,759 | 13,775 |
| Number of nodes | 998 | 982 |

## Graph Rebuilding

The graph drawn after data pruning is depicted in Figure 5. This graph was built after applying the Jaccard Similarity algorithm

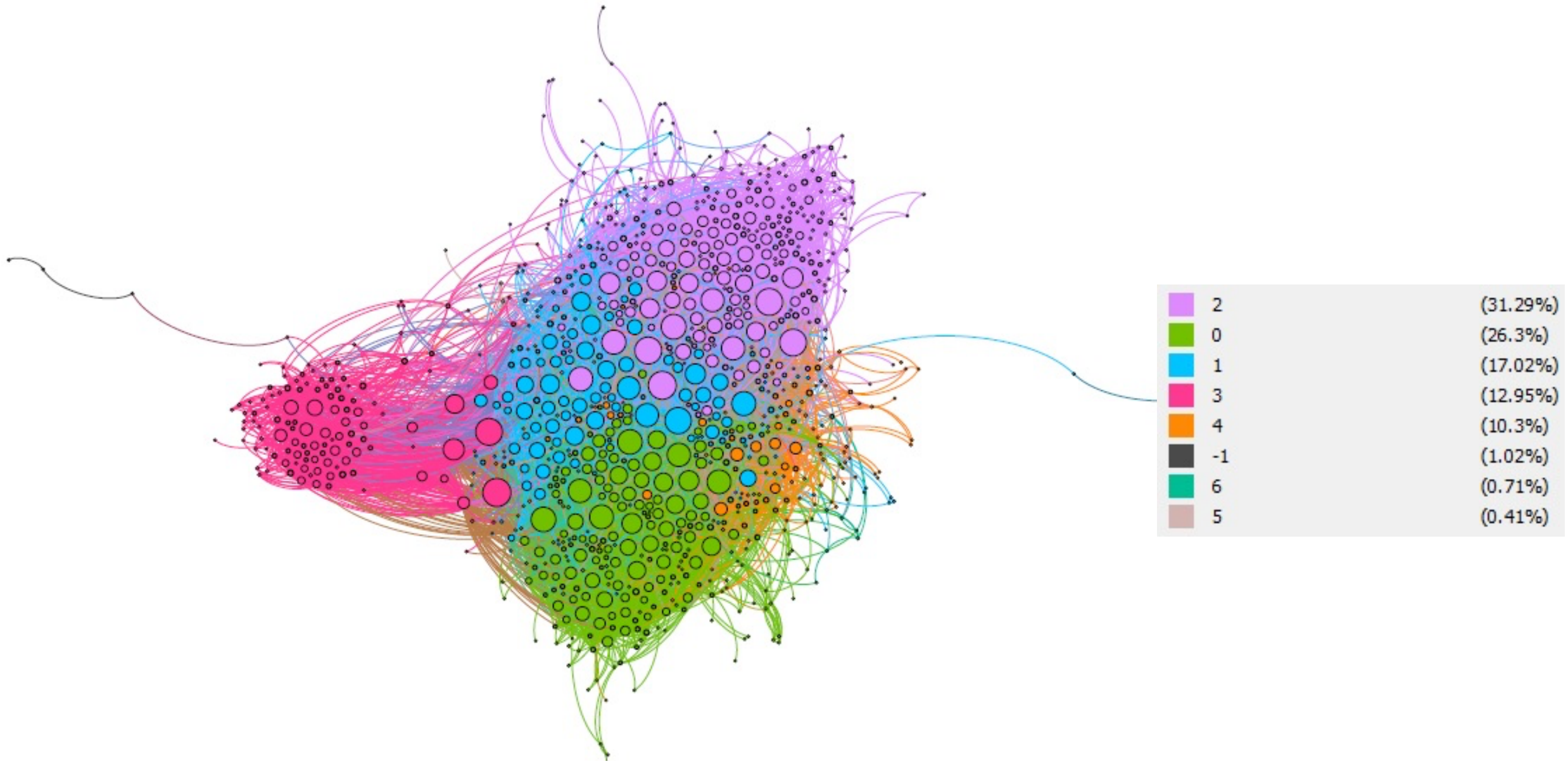

**Figure 5.** A graph is drawn based on post-pruning prescription medicines

The pruned graph results from the second stage of pruning operation in which medicines are divided into eight categories. Class 1 corresponds to noisy data. The algorithm in this step is implemented using the jenkspy library. The clustering of the medicines and the number of medicines in each cluster is shown in Figure 6.

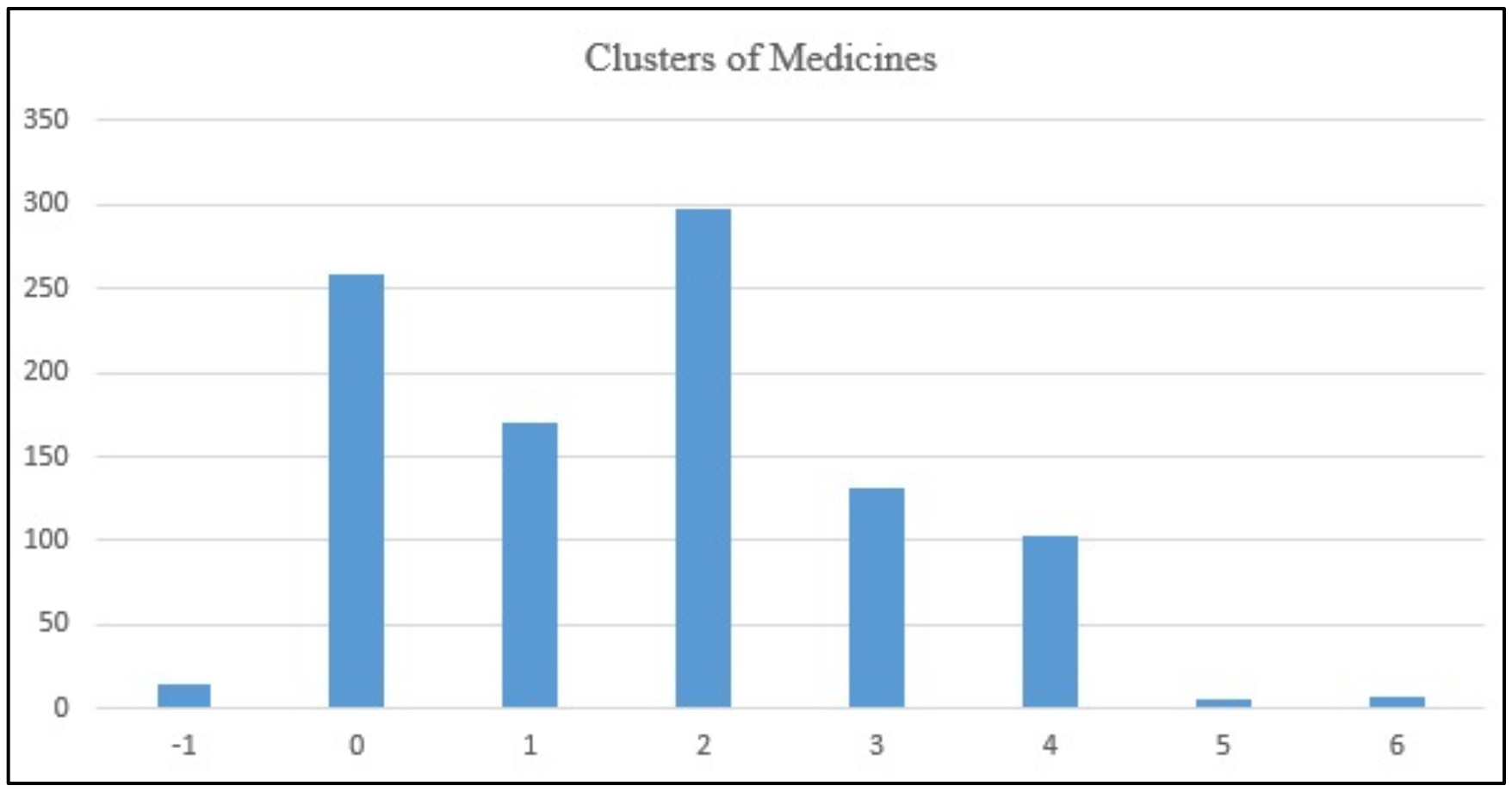

**Figure 6.** Medicine classification

**ATC Matching and Clustering**

The anatomical therapeutic chemical (ATC) classification system has been used to match the medicines to produce an enriched recommendation in this step. The relevant ATC codes have been mapped to equivalent prescription medicines. Finally, medicine data clustering was performed in the final step, using the DBSCAN and Louvain algorithms. Six clusters were produced as the final product in which Class 1 represents the noisy data.

For example, clusters 5 and 6 are presented in Tables 4 and 5. Class 5 contains two medicines: "Iron sucrose 20 mgfe/ml 5ml AMP" and "Erythropoietin Recombinant hu 4000 IU/VIAL", which is an example of an output analysis of this class. Both of these medicines are anemia treatment products used to treat iron deficiency in patients. Class 6 medicines include "prednisolone acetate 1% oph drop", "atropine sulfate 0.5 percent 10ml oph drop", and "Timolol maleate 0.5 percent 5ml oph drops" all class S (sensory organs) medicines. Eyes, ears, and appendages are among them. Therefore, the medicines in this class are similar.

**Table 4.** Clustering output based on prescription medicines

| Id | Class | Medicine |
|---|---|---|
| 11 | 6 | ACETAZOLAMIDE 250MG TAB |
| 118 | 6 | ATROPINE SULFATE 0.5% 10ML OPH DROP |
| 1046 | 6 | PREDNISOLONE ACETATE 1% OPH DROP |
| 1226 | 6 | TIMOLOL MALEATE 0.5% 5ML OPH DROP |
| 1901 | 6 | LATANOPROST 50MCG/ML OPH DROP |
| 11375 | 5 | FERROUS GLYCINE SULPHATE CAP |
| 4003 | 5 | IRON SUCROSE 20 MGFE/ML 5ML AMP |
| 4486 | 5 | ERYTHROPOIETIN RECOMBINANT HU 4000 IU/VIAL |
| 9532 | 5 | MEBEVERINE 200MG ER CAP |

**Table 5.** Class #6 medicine clustering output based on ATC code

| Id | Class | Medicine | ATC Code |
|---|---|---|---|
| 11 | 6 | ACETAZOLAMIDE 250MG TAB | S01EC01 |
| 118 | 6 | ATROPINE SULFATE 0.5% 10ML OPH DROP | S01FA01 |
| 1046 | 6 | PREDNISOLONE ACETATE 1% OPH DROP | S01BA04, S01CB02<br>S02BA03, S03BA02 |
| 1226 | 6 | TIMOLOL MALEATE 0.5% 5ML OPH DROP | S01ED01 |
| 1901 | 6 | LATANOPROST 50MCG/ML OPH DROP | S01EE01 |

Table 6 shows the medicine class #2 as a blood pressure class.

**Table 6**. Sample output of clustering blood pressure medications based on ATC Code

| Id | Class | Medicine | ATC Code |
|---|---|---|---|
| 1045 | 2 | prednisolone 5mg tab | C05AA04 |
| 1067 | 2 | propranolol hcl 10mg tab | C07AA05 |
| 2319 | 2 | metoprolol tartrate *50mg tab* | C07AB02 |
| 2475 | 2 | carvedilol 6.25mg tab | C07AG02 |
| 1822 | 2 | losartan potassium 25mg tab | C09CA01 |
| 1472 | 2 | diltiazem hcl sr 120mg tab | C05AE03<br>C08DB01 |

In this research, the results have been evaluated by an expert person in the medical field and for that, the following steps have been carried out:

**Step1:**

A subset of medicines was randomly selected from class #2 (hypertension medicine class).

**Step2:**

A human expert performed the medicine labeling by considering the ATC Code values of the medicines.

**Step3:**

And finally, the real labels which were provided by the expert have been compared with the labels of medicine which were recommended by the system. If the medicine label was correctly identified by our proposed methods, the human expert marked a 1 score and if the label of medicine was not recognized correctly it would be marked as a 0 score which means a wrong diagnosis.

**Table 7**. Evaluation results of hypertensive medicines output based on expert opinion and ATC Code

| # | Id | Medicine | Tag | ATC Code |
|---|---|---|---|---|
| 1 | 1045 | prednisolone 5mg tab | 1 | C05AA04 |
| 2 | 1067 | propranolol hcl 10mg tab | 1 | C07AA05 |
| 3 | 2319 | metoprolol tartrate *50mg tab* | 1 | C07AB02 |
| 4 | 122 | azathioprine 50mg tab | 0 | L04AX01 |
| 5 | 2475 | carvedilol 6.25mg tab | 1 | C07AG02 |
| 6 | 1822 | losartan potassium 25mg tab | 1 | C09CA01 |
| 7 | 1472 | diltiazem hcl sr 120mg tab | 1 | C05AE03<br>C08DB01 |
| 8 | 206 | captopril 50mg tab | 1 | C09AA01 |
| 9 | 447 | diltiazem hcl 60mg tab | 1 | C05AE03 |
| 10 | 205 | captopril 25mg tab | 1 | C09AA01 |
| 11 | 1888 | buspirone 5mg tab | 1 | N05BE01 |
| 12 | 70 | amlodipine 5mg tab | 1 | C08CA01 |
| 13 | 114 | atenolol 100mg tab | 1 | C07AB03 |
| 14 | 115 | atenolol 100mg tab | 1 | C07AB03 |
| 15 | 1068 | propranolol hcl 40mg tab | 1 | C07AA05 |
| 16 | 1169 | spironolactone 25mg tab | 1 | C03DA01 |
| 17 | 1252 | triamterene-h tab | 1 | C03DB02 |
| 18 | 1472 | diltiazem hcl sr 120mg tab | 1 | C08DB01 |
| 19 | 1520 | gemfibrozil 450mg tab | 1 | C10AB04 |
| 20 | 1913 | celecoxib 100mg cap | 1 | C08CA51 |
| 21 | 1924 | carvedilol 12.5mg tab | 1 | C07AG02 |
| 22 | 1925 | carvedilol 25mg tab | 1 | C07AG02 |
| 23 | 1933 | losartan potassium 50mg tab | 1 | C09CA01 |
| 24 | 2319 | metoprolol tartrate 50mg tab | 1 | C07AB02 |
| 25 | 6788 | rosuvastatin calcium 20mg tab | 1 | C10AA07 |
| 26 | 6459 | nicorandil 10mg tab | 1 | C01DX16 |
| 27 | 7066 | propranolol hcl 20mg tab | 1 | C07AA05 |
| 28 | 51 | amantadine hcl 100mg cap | 1 | N04BB01 |
| 29 | 1009 | pimozide 4mg tab | 1 | N05AG02 |
| 30 | 68 | amitriptyline 25mg tab | 1 | N06AA09 |

In Table 7, evaluation of class #2 by the human expert is shown as the "Tag" column. Based on the results obtained from the expert evaluation, all samples, except one of them, were labeled correctly.

# 4. Discussion and Subjective Comparison

Each recommender system, in general, focuses on a single topic. Thus, the design steps, recommendation techniques, and, finally, implementation are completed following it. This is done so that the recommender system has the best performance in that particular domain. The general design steps for recommender systems are the same regardless of the different applications and scope. To subjectively compare medicine recommender systems in this study to other previously introduced recommender systems, it can be said that recommender system implementation is divided into two steps:

1- Collecting a dataset for building the recommender engine

2- Mining data and associations between primary datasets to generate recommendations.

The components of a recommender system are examined and compared to those of other similar systems in the table below.

Table 8. A subjective comparison of other recommender systems and the introduced medicine recommender system

| Parameters / Papers | Our proposed medicine recommender system | The intelligent medicine recommender system [25] | The medicine recommender system in [26] | The doctor recommender system [14]–[16] | Diabetic patients' medicine recommender system [21] |
|---|---|---|---|---|---|
| **Scope** | Field of blood pressure medications | Field of medicine recommendation | Field of medicine recommendation | Field of doctor recommendation | Lot of diabetic patients' medicines |
| **Data type** | Prescriptions collected from all over the country | Age, gender, blood pressure, cholesterol, consumed medicines | Information such as name, name of medicine, name of disease; and the name of the medicine | Personal information on patient's health including consumed medicines, medication dose, side-effects of surgeries, and other health indicators such as blood pressure, weight, test result, the hospital in which the patient is hospitalized | Medicines consumed by the user, diet, exercise done by the patient, blood insulin levels, blood pressure, age, glucose levels, and BMI |
| **Labeled/unlabeled data** | Unlabeled data | Labeled data | Labeled data | Labeled data | Labeled data |
| **Preprocessing** | Yes | Yes | Yes | No | No |
| **Pruning** | Pruning for the removal of stop medicines, i.e., general medications, as proposed by the doctor | Pruning for placing medicines over a specific interval and equating the effectiveness degree of medicines | Pruning is not done. | Pruning is not done. | Pruning is not done. |
| **Similarity measurement techniques** | Using the Jaccard similarity technique to calculate the degree to which two medicines simultaneously present in a prescription are similar | Using the chi-square method to determine the degree to which nominal variables and medicine variables, except for gender, are associated | Similarity measurement techniques are not employed. | Calculating user similarities depending on the conditions and diseases as well as the similarity score using the semantic similarity technique | Measuring user similarities using Pearson's correlation coefficient |
| **Recommendation model technique** | Association [rule mining] algorithms and clustering algorithms such as Louvain and DBSCAN | SVM algorithm | SVM algorithm | Calculating patient similarities and creating the doctor-hospital scoring list | User-Based Collaborative Filtering |
| **Visualization** | Graphical representation of pre-and post- pruning graph using the Gephi software to illustrate the association between medicines, pre-and post-pruning node-edge count diagram, medicine classification based on the Fisher-Jenks algorithm, and medicine clustering and the number of medicines in each class | Drawing graphical representations using Python libraries, medicine dispersion diagram, medicine-gender diagram, the learning trial-and-error diagram, and the visualized model running accuracy and time diagram | Visualization is not used. | Visualization is not used. | Visualization is not used. |
| **The medical standard for medical improvement** | Utilizing the anatomical therapeutic chemical (ATC) classification system known as the ATC Code to analyze the recommendations | Discussing the mistaken-check mechanism, and at the same time, avoiding using medical standards to improve responses | No medical standards used | No medical standards used | No medical standards used |
| **Output** | Recommending what medicines can be prescribed together | Offering medicine recommendations | Offering medicine recommendations | Doctor's and hospital's recommendations offered to the patient | Diet and exercise for diabetic patients |

## 5. Conclusions and Future Work

In summary, the proposed method in this research is divided into two steps. The first step involves extracting the rules and relations between drugs by using the association rule mining algorithms that lead to the extraction of rules and relations between the drugs, and the second step involves the extraction of the graph and clusters to present an enriched recommendation via ATC code. The use of the ATC code which is a classification system maintained by the World Health Organization (WHO) can provide high quality and acceptable results to prescribe blood pressure drugs which are usually used together.

Unlike other research studies, in this work, the framework proposed is entirely based on unsupervised learning methods and so it is more likely to be used in real-world applications as medical data are generally unlabeled, and supervised learning models cannot be directly built using these types of data. In the end, the output was reviewed, checked and confirmed by an expert person.

In future work, in addition to historical medicine datasets, clinical data can be utilized to enhance further the proposed approach's accuracy. It is possible to recommend medicines by collecting a database of patients' clinical symptoms resulting from clinical questionnaires designed by the medical team, including personal information such as age, height, weight (BMI), daily activity, patient history, underlying diseases, and consumable medicines. This is only possible with the assistance of a medical team. Combining these two medical and clinical datasets creates a system with a high confidence score and a lower error rate. The work presented here can also be applied to datasets from other countries to design a medicine recommender system tailored to that country's medical and patient conditions.